\documentclass{dcdsc} 
\usepackage{bbding}
\usepackage{subcaption}
\allowdisplaybreaks

\newcommand{\doititle}[1]{#1}

\def\currentvolume{xx}

\def\currentyear{20xx}
\def\ppages{xxx-xxx}
\def\DOI{10.3934/dcdsc.20xxxxx}

\theoremstyle{plain}

\theoremstyle{definition}

\title{From Basins to Safe Sets: A Machine Learning Perspective on Chaotic Dynamics}

\author[a]{\scshape David Valle\textsuperscript{\href{mailto:david.valle@urjc.es}{\Envelope}}}
\author[a]{\scshape Alexandre Wagemakers\textsuperscript{\href{mailto:alexandre.wagemakers@urjc.es}{\Envelope}}}
\author[*,a,b]{\scshape Miguel A.F. Sanjuán\textsuperscript{\href{mailto:miguel.sanjuan@urjc.es}{\Envelope}}}

\begin{document}

\thispagestyle{cover}
\FirstpageLayout

\maketitle
\medskip

\noindent\text{Affiliations are at the bottom of this page.}
\medskip

\FirstpageAffiliations{%
  \textsuperscript{a}Nonlinear Dynamics, Chaos and Complex Systems Group, Departamento de Física, Universidad Rey Juan Carlos, Tulipán s/n, 28933 Móstoles, Madrid, Spain.\\
  \textsuperscript{b}Royal Academy of Sciences of Spain,
  Valverde 22, 28004 Madrid, Spain.
}


\bigskip
\bigskip

\begin{abstract}
The study of chaos has long relied on computationally intensive methods to quantify unpredictability and design control strategies. Recent advances in machine learning, from convolutional neural networks to transformer architectures, provide new ways to analyze complex phase space structures and enable real time action in chaotic dynamics. In this perspective article, we highlight how data driven approaches can accelerate classical tasks such as estimating basin characterization metrics, or partial control of transient chaos, while opening new possibilities for scalable and robust interventions in chaotic systems. In recent studies, convolutional networks have reproduced classical basin metrics with negligible bias and low computational cost, while transformer based surrogates have computed accurate safety functions within seconds, bypassing the recursive procedures required by traditional methods. We discuss current opportunities, remaining challenges, and future directions at the intersection of nonlinear dynamics and artificial intelligence.
\end{abstract}

\bigskip
\keywords{Chaos control, machine learning, neural networks, basins of attraction, transient chaos.}


\ReceivedDate{xxxx xx, 20xx}
\AcceptedDate{xxxx xx, 20xx}
\PublishedDate{xxxx xx, 20xx}

\Significance{%
Chaos plays a central role in many natural and engineered systems, yet its analysis and control are often limited by high computational costs. This work shows how modern machine learning techniques can accelerate classical tasks in nonlinear dynamics, such as the characterization of basins of attraction and the control of transient chaos. By combining established dynamical systems theory with data-driven models, the proposed approaches enable faster analysis and open the door to real-time applications in complex systems. These results are relevant for a broad scientific audience interested in predictability, robustness, and control across physics, engineering, and applied mathematics.
}

\FirstPageRightBlockBottom{%
  \textbf{Author contributions}:
    David Valle designed the research and performed numerical experiments;
    Alexandre Wagemakers contributed to the theoretical analysis and interpretation of results;
    Miguel A.F. Sanjuán supervised the project and contributed to the conceptual framework;
    and all authors wrote and revised the manuscript.\\
  \textbf{Competing interests}: The authors declare no competing interests.\\
  \textsuperscript{1}\textbf{Co-first authors}\\
  \CorrespondingAuthor{miguel.sanjuan@urjc.es} \\
 \textbf{Handling Editor}: To be assigned\\
%
  \Subjclass{Primary: 37D45, 68T07; Secondary: 37N30, 93C41.}
}
\FirstPageRightBlock




\section*{Introduction}

Chaos, long recognized as a hallmark of complexity in dynamical systems, continues to challenge our ability to predict and control physical, biological, and engineered processes~\cite{Strogatz2015}. At its core, chaos emerges from deterministic equations, yet it exhibits extreme sensitivity to initial conditions and complex phase space structures~\cite{Lorenz1963}. This peculiar coexistence of determinism and unpredictability has profound implications for understanding multistability, resilience, and failure in a wide range of systems, such as climate dynamics, power grids, and ecosystems~\cite{Feudel1997}.

\begin{figure}[t]
  \centering
  \begin{subfigure}[a]{0.41\textwidth}
    \centering
    \includegraphics[width=\linewidth]{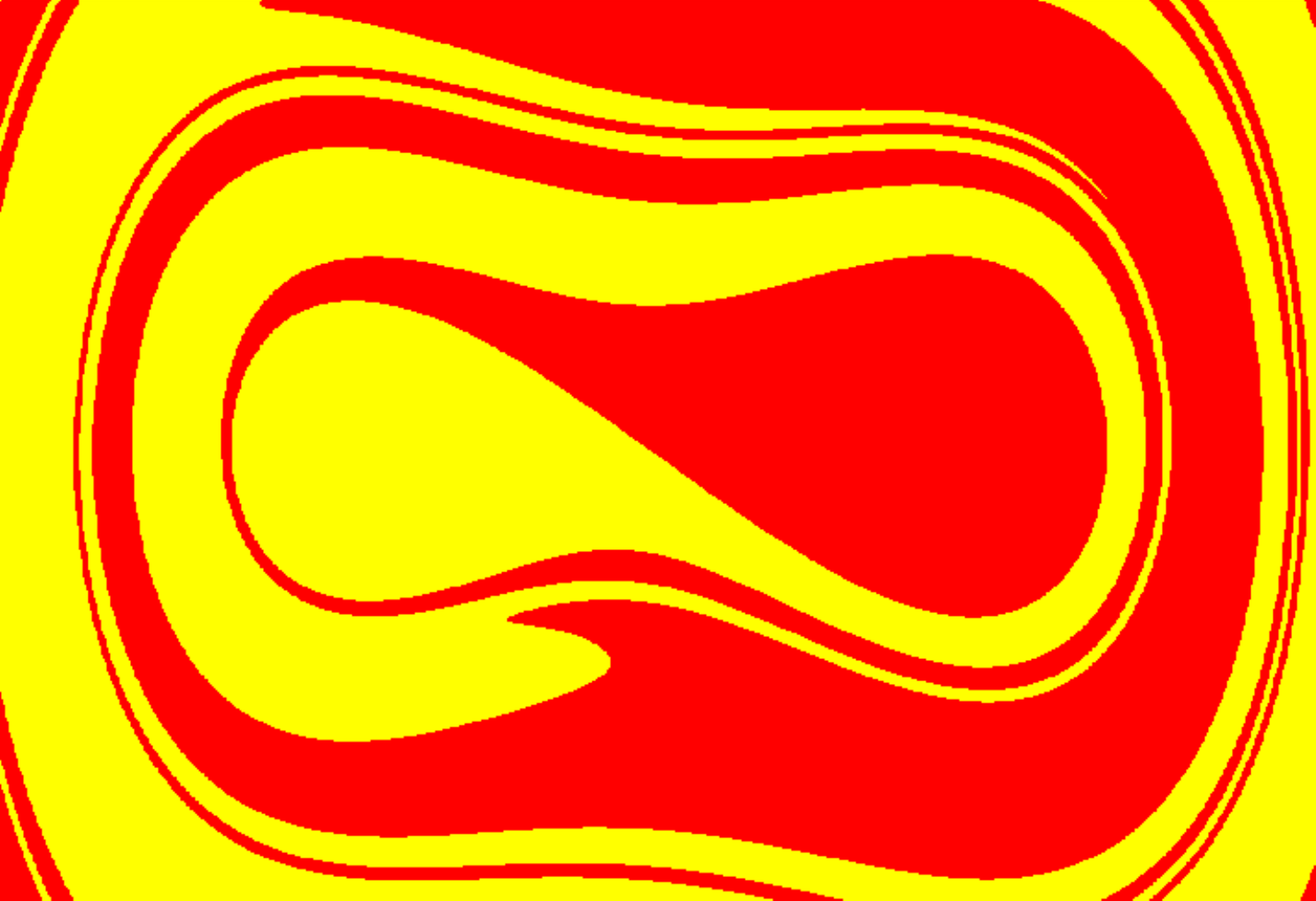}
    \subcaption{}\label{fig:basin-smooth}
  \end{subfigure}
  \hfill
  \begin{subfigure}[a]{0.41\textwidth}
    \centering
    \includegraphics[width=\linewidth]{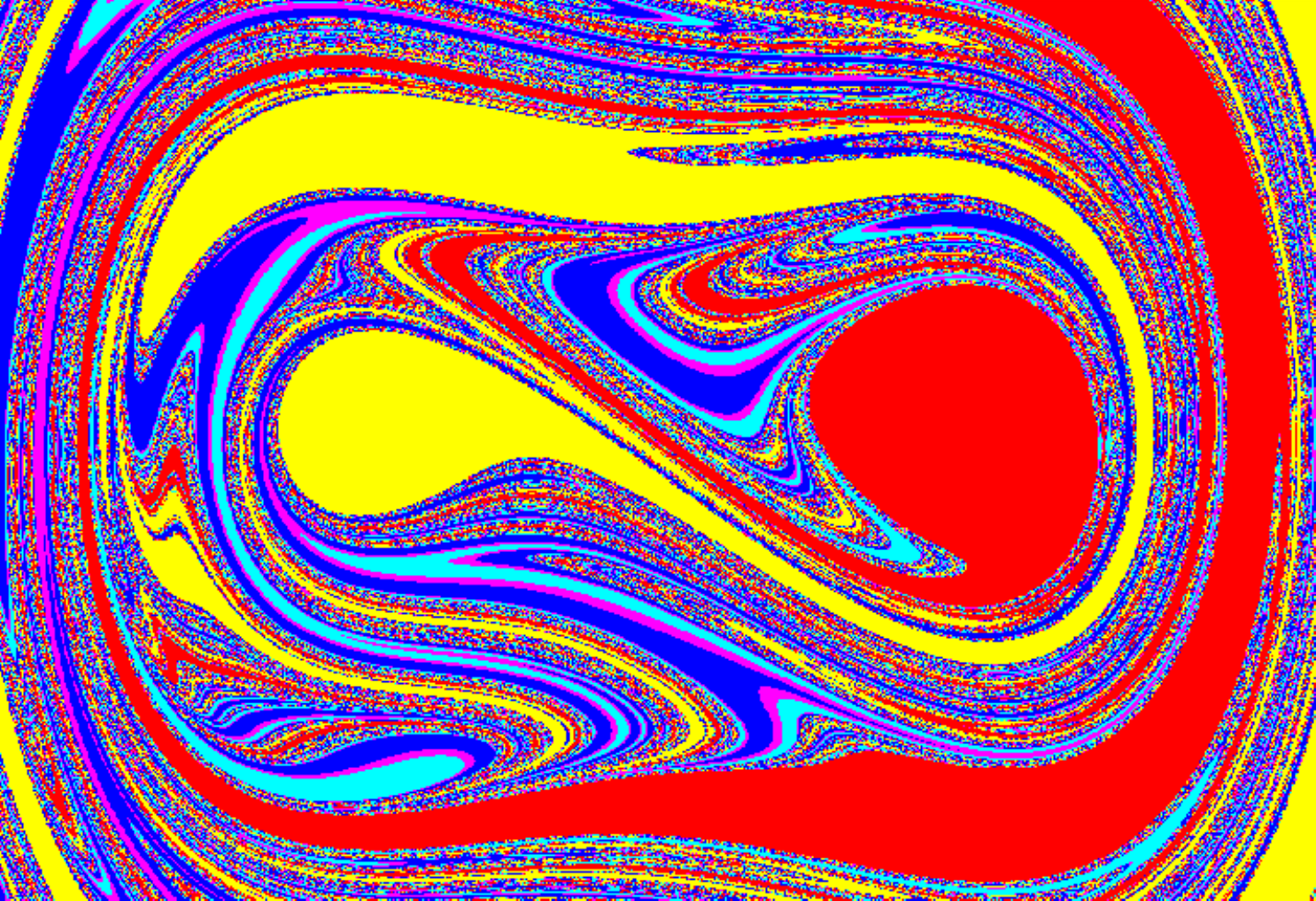}
    \subcaption{}\label{fig:basin-fractal}
  \end{subfigure}
  \caption{Basins of attraction in a two dimensional dynamical system. Each pixel corresponds to an initial condition, and the color denotes its asymptotic attractor. (a) Example with smooth basin boundaries, where nearby initial conditions converge to the same attractor, indicating high predictability and low basin entropy. (b) Example with fractal basin boundaries, where arbitrarily close initial conditions may converge to different attractors, exemplifying sensitive dependence on initial conditions, increased unpredictability and higher basin entropy.}
  \label{fig:basins-comparison}
\end{figure}

One of the most striking manifestations of chaos lies in the structure of basins of attraction, which partition the phase space into regions corresponding to different asymptotic states~\cite{Nusse1997}. The boundaries of these basins may exhibit fractal geometry~\cite{McDonald1985}, high basin entropy~\cite{Daza2016}, and the Wada property~\cite{Aguirre2009}, reflecting the deep unpredictability embedded in the system. This contrast is illustrated in Fig.~\ref{fig:basins-comparison}, which compares a basin with smooth boundaries, associated with higher predictability, to one with highly fractal boundaries, where unpredictability is maximized.

\Endparasplit

Quantifying these properties relies on numerical methods, such as box-counting for estimating fractal dimension~\cite{McDonald1985}, entropy based techniques for unpredictability analysis~\cite{Daza2016}, and a variety of algorithms for verifying the Wada property. The latter includes the Nusse–Yorke method~\cite{Nusse1996}, the grid method~\cite{Daza2015}, the saddle-straddle method~\cite{Wagemakers2020}, and the merging method~\cite{Daza2018}, each one with specific advantages and limitations. The choice of the method depends on the available information about the system and the desired level of confidence in the result, with higher reliability often requiring increased computational effort. A concise overview of these techniques can be found in~\cite{Wagemakers2021}. While rigorous, these classical methods are computationally expensive, particularly when applied to large parameter spaces, high dimensional systems, or scenarios demanding real time analysis.

The unpredictability of the final state or the chaotic dynamics of the time series motivated approaches to control such uncertainties. Traditional chaos control methods, such as the Ott–Grebogi–Yorke (OGY) method~\cite{Ott1990}, aim to stabilize chaotic trajectories by applying small, precisely timed perturbations to guide the system toward unstable periodic orbits embedded within the chaotic attractor. These techniques have proven effective in many low dimensional systems where a suitable target orbit is known and can be accessed through local linearization~\cite{Ott1990}. However, in the context of transient chaos~\cite{Lai2011}, where chaotic motion eventually collapses into regular dynamics, the objective is often not to stabilize a trajectory indefinitely, but to prolong the chaotic regime or guide the system toward desirable asymptotic states while avoiding escape into unwanted attractors~\cite{Dhamala1999PRE}. This fundamental shift in control objectives renders classical methods less effective and motivates the development of alternative frameworks.

\begin{figure}[t]
  \centering
  \includegraphics[width=1\linewidth]{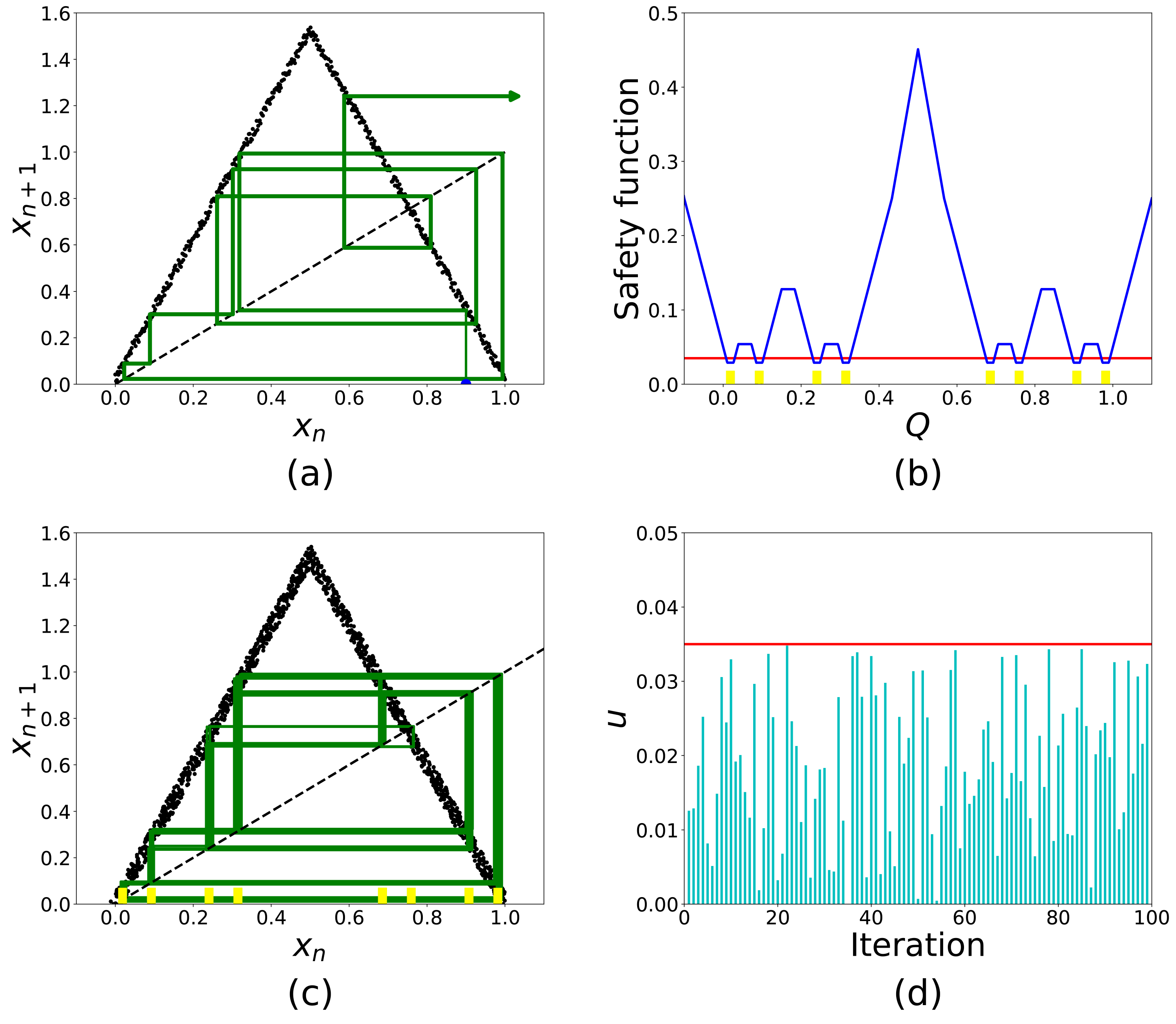}
  \caption{Safety function and safe sets. 
  (a) Uncontrolled trajectory (green) escaping the region $Q=[0,1]$. 
  (b) Safety function $U_\infty(x)$ (blue) with admissible control bound $u$ (red), the safe set $S(u)=\{x\in Q: U_\infty(x)\le u\}$ is the yellow region below this bound. 
  (c) Controlled trajectory (green), starting from the same initial condition, confined within $S(u)$. 
  (d) Control signal $u_n$ (cyan), bounded by $u$ (red). 
 This qualitative illustration shows how partial control confines chaotic trajectories with minimal interventions.}
  \label{fig:safety-example}
\end{figure}

One such framework is partial control~\cite{Sabuco2012}, which tackles the problem of maintaining trajectories within a compact target region $Q$ of the phase space, typically containing the chaotic saddle, by applying interventions weaker than the disturbances driving the system. Central to this approach is the safety function $U_\infty(q)$, which assigns to each state the minimal control required to prevent escape, and the corresponding safe set $S(u)=\{q\in Q: U_\infty(q)\le u\}$, representing all states that can be confined under a given control bound $u$~\cite{Capeans2019}. Figure~\ref{fig:safety-example} illustrates these ideas: an uncontrolled trajectory leaves the region (panel a), but by evaluating the safety function (panel b), one can identify the safe set (yellow) associated to a given control threshold. Within this set, trajectories can be confined using bounded inputs (panels c and d). While conceptually powerful, computing safety functions requires recursive, high resolution sweeps of the state space, which becomes prohibitive in noisy or high dimensional contexts~\cite{Capeans2017,Capeans2022}.

The growing gap between theoretical robustness and computational feasibility has underscored the need for new approaches. Advances in machine learning, combined with abundant dynamical data and modern computational resources, offer timely opportunities to revisit longstanding challenges in chaos theory~\cite{Brunton2016,Pathak2018}. As we discuss next, data driven strategies are reshaping both the analysis and control of chaotic dynamics.

\section*{State of the Art}

The study of chaotic dynamics has historically relied on classical numerical techniques to quantify unpredictability and control behavior. Metrics such as the fractal dimension of basin boundaries~\cite{McDonald1985}, basin entropy and boundary basin entropy~\cite{Daza2016}, and detection of the Wada property~\cite{Aguirre2009} remain central to characterizing the complexity of basins of attraction. These tools quantify sensitivity to initial conditions of a final state. Yet, despite their theoretical rigor, classical algorithms, such as box-counting for fractal dimension or Wada detection using grids, suffer from high computational costs and limited scalability, particularly in high resolution parameter studies or real time settings~\cite{Valle2022,Valle2024}. 

Recent advances in Machine Learning (ML), particularly deep learning, have begun to address these limitations by learning from data the nonlinear relationships that underlie basin geometry. Convolutional neural networks have been trained on basin images to estimate fractal dimension, basin entropy, or boundary basin entropy with accuracy comparable to classical algorithms but at substantially lower computational cost~\cite{Valle2022,Valle2024}. In basin characterization, CNN surrogates learn the nonlinear mapping from basin images to metrics, replacing repeated box-counting or Monte Carlo sweeps with a single forward pass. For instance, in the estimation of fractal dimension, CNNs achieve errors of order $10^{-2}$ while reducing compute time from several seconds to well below one second per basin, yielding roughly an order of magnitude improvement on comparable hardware with negligible setup cost once trained~\cite{Valle2022,Valle2024}.

Classical control strategies such as the Ott–Grebogi–Yorke (OGY) method~\cite{Ott1990} stabilize chaotic trajectories by small targeted perturbations, but these techniques are limited to accessible unstable periodic orbits. In transient chaos~\cite{Lai2011}, where chaotic motion eventually collapses into regular dynamics, the partial control framework~\cite{Sabuco2012,Capeans2019} offers an alternative: computing safety functions and safe sets to determine where bounded control succeeds. However, these algorithms require recursive high resolution sweeps of the state space and become prohibitively expensive in noisy or high dimensional systems~\cite{Capeans2017,Capeans2022}.  

Machine learning has recently enabled more efficient surrogates for these control computations. Transformer-based models, originally developed for sequence learning, have been adapted to approximate safety functions directly from trajectory data, bypassing the need for recursive evaluation and enabling real time, model free control~\cite{Valle2025CNSNS,Valle2025EPJST}. In practice, transformer surrogates predict $U_\infty$ directly from short trajectory segments, omitting the nested recursions of the classical algorithm that become computationally prohibitive as dimensionality increases. On representative one dimensional maps, the mean squared errors are of order $10^{-4}$, and the prediction times remain similar to those of classical computations, indicating that the method is computationally reasonable for practical use~\cite{Valle2025CNSNS}. The main advantage lies in obtaining accurate safety functions directly from sampled trajectories without requiring full knowledge of the underlying dynamics. This approach, beyond its applicability in one dimensional maps, is expected to scale favorably to higher dimensional systems, as the computational complexity of transformer training grows more slowly than that of classical recursive algorithms~\cite{Valle2025EPJST}.

\section*{Open Problems}

Despite recent advances, several key challenges remain for the integration of machine learning and nonlinear dynamics. A first open direction concerns the automated identification of complex phase space structures such as riddled basins or KAM islands~\cite{Ott1993PRL, Lai1995PRE, Broer1996CMP}. Current methods still rely on visual inspection or indirect geometric indicators, and building reliable classifiers capable of detecting these fine structures across systems and parameter regimes remains an unresolved task.

A second major challenge lies in the extension of partial control and safety function computation to higher dimensional systems. Classical recursive algorithms scale poorly with dimension, and even current ML surrogates have been trained mostly on low dimensional maps. Developing architectures and sampling strategies that generalize efficiently in dimension while preserving control guarantees would substantially expand the applicability of these techniques.

Another open problem is uncertainty quantification and interpretability in ML surrogates. Quantifying confidence in predicted basin metrics or safety functions is essential for their reliable use in scientific or engineering applications. Bayesian or ensemble based uncertainty estimation, combined with feature attribution or saliency analysis, could help identify which regions of phase space drive model predictions and expose potential failure modes.

Finally, there is a need for standardized benchmarks and hybrid validation pipelines. Shared datasets of basins, trajectories, and safety functions would enable consistent evaluation of ML approaches against classical baselines.

\section*{Outlook and Future Directions}

We envision a future where machine learning methods not only complement but also expand the classical toolbox of chaos analysis and control. Recent progress shows that ML can extract actionable patterns from data at scales and speeds unattainable for traditional approaches. Yet its full potential will only be realized through careful integration with nonlinear dynamics. Below, we outline three directions where this integration could reshape the field.

\textbf{(i) Characterization of basins of attraction.} ML has already proven capable of predicting key unpredictability metrics such as fractal dimension, basin entropy, boundary basin entropy, and the presence of the Wada property from basin data~\cite{Valle2022,Valle2024}. Building on this foundation, priorities include the automated estimation of intricate structures such as riddled basins and KAM islands, the extension of these analyses to additional global metrics such as basin stability~\cite{Menck2013,Leng2016}, improved efficiency for real time and high dimensional studies through super resolution, generative models, or hybrid ML classical pipelines, and the incorporation of explainability mechanisms that link predictions to geometric features to improve interpretability and trust.

\textbf{(ii) Control of transient chaos.} Partial control based on safety functions and safe sets remains limited by high computational demands~\cite{Capeans2019,Capeans2022}. Transformer-based models that approximate safety functions directly from trajectories provide a promising, real time alternative~\cite{Valle2025CNSNS,Valle2025EPJST}. A pressing challenge is scaling these methods to higher dimensional systems, where classical computations quickly become intractable. Dimensionality and scalability are central issues. Classical partial control scales poorly because of grid growth and nested optimizations, while ML surrogates mitigate this growth through amortization, the costly search is performed once during training and inference becomes grid free. For higher dimensional states, dimensionality reduction and domain adaptation become natural complements, learning low dimensional embeddings through autoencoders, diffusion maps, or Koopman based coordinates where $U_\infty$ varies smoothly, exploiting symmetries, and adapting models across parameter regimes via fine tuning or conditional inputs. These strategies reduce sample complexity and allow safety function surrogates to operate efficiently in spaces where classical meshes are infeasible. Another promising avenue is to design and optimize safety functions under alternative criteria: instead of minimizing the maximum control required to prevent escape, one could minimize the average control applied over several iterations, reducing intervention costs while maintaining confinement. Coupling transformer-based approximations with reinforcement learning agents could also enable adaptive, context aware controllers that self tune under noise, parameter drift, or unexpected disturbances, shifting from static maps to dynamic, resilient strategies. Such approaches could prove transformative in applications ranging from climate models to power grids or neural systems.

\textbf{(iii) Interpretability and hybrid models.} A persistent challenge for ML in chaotic dynamics is its black-box nature, which limits trust and scientific insight. Embedding physical constraints into learning architectures, such as physics-informed neural networks (PINNs)~\cite{Raissi2019JCP}, constrained optimization layers~\cite{Amos2017OptNet}, or physics-guided regularization~\cite{Karpatne2022PGNN}, can enforce invariants, symmetries, or conservation laws, ensuring predictions remain consistent with known dynamics. Symbolic regression and sparse equation discovery techniques~\cite{Brunton2016} offer another route, translating data driven patterns into parsimonious governing equations that reveal effective laws behind complex transients. Uncertainty aware models, for instance Bayesian neural networks~\cite{Gal2016Dropout} or deep ensembles~\cite{Lakshminarayanan2017Ensembles}, could provide confidence estimates alongside predictions of basin metrics or safety functions, establishing principled baselines against classical methods. Visualization methods such as saliency maps~\cite{Simonyan2014Saliency} or feature attribution analyses~\cite{Ribeiro2016LIME}, though originally developed for computer vision, could be adapted to identify which features of basins or trajectories most influence model outputs. The scientific objective for interpretability should go beyond generic saliency to identify which phase space features drive predictions of $U_\infty$ and $S(u)$. For basins, model attributions could be linked to boundary filaments and Wada junctions to quantify which microstructures dominate uncertainty. For one dimensional maps, attributions of $U_\infty(y)$ to local slope, distance to escape sets, and noise bounds could be visualized directly on $f(y)$ and its disturbed images. Such analyses would reveal control relevant structures, such as narrow bottlenecks in $S(u)$, and guide hybrid refinements that allocate classical verification exactly where the surrogate is least certain. Ultimately, hybrid pipelines that combine the speed of ML with the rigor of classical validation will be essential to ensure both scalability and reliability in chaos research.

More broadly, progress will require a collaborative ecosystem where ML models are trained and validated on shared benchmark datasets of trajectories, basins, and safety functions, covering diverse systems and noise levels. Such resources would enhance reproducibility and allow meaningful comparisons across methods.

In summary, ML has the potential to become a standard component of the chaos research toolkit: fast, scalable, and interpretable. Realizing this vision requires advances in generalization, interpretability, and robustness, together with rigorous benchmarking against classical techniques. By embracing this integration, the community can move beyond describing unpredictability toward actively steering complex dynamics across physics, engineering, and beyond.

\section*{Conclusion}

In this perspective we have reviewed how recent machine learning methods can complement the classical approaches used in the analysis and control of chaotic dynamics. Our intention is not to present ML as a substitute or a pioneering solution, but rather to place current advances in context and to highlight specific areas where these tools may offer practical advantages.

Regarding analysis, previous studies have shown that convolutional neural networks can speed up the estimation of unpredictability indicators from basin data. Building on these results, future work may extend to the detection of more complex structures or to efficiency gains through hybrid pipelines that combine ML with classical algorithms. In the field of control, transformer based models that approximate safety functions from trajectory data open the door to faster and more flexible implementations of partial control. These approaches should be seen as accelerators that work alongside established methods, rather than as replacements.

Looking ahead, three priorities emerge. First, to clarify the scope of application: ML can be especially useful in large parameter sweeps, online analysis, or noisy datasets, while classical methods remain essential when rigorous guarantees are required. Second, to ensure reliability: fast predictions should be combined with uncertainty estimates and systematic comparisons with traditional baselines. Third, to improve comparability: building shared datasets of basins, trajectories, and safety functions would allow consistent evaluation across systems.

In summary, our contribution is to provide an integrated view and a practical direction: to use ML where it alleviates computational bottlenecks, to validate results through classical methods, and to work toward common benchmarks. With this approach, ML can make classical chaos theory more effective and applicable in complex or time sensitive contexts.

\section*{Acknowledgments}
This work has been financially supported by the Spanish State Research Agency (AEI) and the European Regional Development Fund (ERDF, EU) under Project No.~PID2023-148160NB-I00 \\(MCIN/AEI/10.13039/501100011033).









\medskip
\medskip

\end{document}